\documentclass[11pt]{amsart}

\usepackage{amsmath, amssymb, graphics}
\usepackage{graphicx}

\newtheorem{thm}{Theorem}[section]

\newtheorem{defn}[thm]{Definition}

\numberwithin{equation}{section}
\newcommand{\mathsym}[1]{{}}

\begin{document}

\title[Perturbation Theory on the NC Plane]{Perturbation Theory on the Non-Commutative Plane with a Singular Potential}

\author{M.\ Nieto}%
\address{Departamento de F\'{\i}sica, Facultad de Ciencias Exactas, Universidad Nacional de La Plata, C.C.\ 67, (1900)  La Plata, Argentina}%
\email{mnieto@fisica.unlp.edu.ar}%

\author{P.A.G.\ Pisani}%
\address{IFLP - CONICET / Departamento de F\'{\i}sica, Facultad de Ciencias Exactas, Universidad Nacional de La Plata, C.C.\ 67, (1900)  La Plata, Argentina}%
\email{pisani@fisica.unlp.edu.ar}%

\author{H.\ Falomir}%
\address{IFLP - CONICET / Departamento de F\'{\i}sica, Facultad de Ciencias Exactas, Universidad Nacional de La Plata, C.C.\ 67, (1900)  La Plata, Argentina}%
\email{falomir@fisica.unlp.edu.ar}%

%\thanks{Work partially supported by CONICET (PIP 01787), ANPCyT (PICT 00909) and UNLP (Proy.~11/X492), Argentina.}%

%\subjclass{}%
%\keywords{}%

%\date{\today}%
\date{November 16, 2010.}%

\maketitle

\begin{abstract}
In this article we study the problem of a non-relativistic particle in the presence of a singular potential in the noncommutative plane. The potential contains a term proportional to $1/R^2$, where $R^2$ is the squared distance to the origin in the noncommutative plane. We find that the spectrum of energies is non analytic in the noncommutativity parameter $\theta$.
\end{abstract}
%\pacs{PACS numbers:12.60.-i,11.30.Cp}
%\date{\today}
\maketitle

%%%%%%%%%%%%%%%%%%%%%%%
\vspace{1cm}

\section{Introduction}

Noncommutative geometry has several motivations in physics \cite{Connes:1996gi,C-M} (see also \cite{Madore:1999bi}
and references therein.) In particular, it has been shown to play an important role in super\-string/M-theory \cite{Witten:1985cc,Seiberg:1999vs} since, in the presence of a Neveu-Schwarz constant background field, low energy string theory reduces to a gauge theory in a noncommutative space. Since then there has been a growing interest in the pecularities of quantum theory on noncommutative spaces.

Research in noncommutative quantum field theories (see the reviews \cite{Douglas:2001ba,Szabo:2001kg}) has motivated the study of the spectral properties of operators defined on noncommutative manifolds (see, e.g., the short review \cite{Vassilevich:2007fq}.)

Central potentials in Quantum Mechanics on non-commutative space have been studied employing perturbation theory on the parameter of noncommutativity. Some examples are \cite{Gamboa:2001qa,Chaichian:2001si,testing,Kochan:2001pz,Adorno:2009yu,Gomes:2009rz}. Delta-function potentials on the non-commutative plane have also been considered in \cite{Yelnikov}.

The aim of the present article is to show how a singular central potential in the noncommutative plane can spoil the analyticity of the spectrum in the noncommutativity parameter $\theta$, therefore ruling out the usual perturbative treatment of noncommutative effects.

We consider a non-relativistic particle on the noncommutative plane subject to a potential consisting on a term proportional to $R^2$ --the square of the noncommutative distance to the origin-- and a singular repulsive term $\alpha/R^2$, with $\alpha\in\mathbb{R}^+$. The former excludes scattering states, so that the (discrete) spectrum of the Hamiltonian corresponds only to bound states. The main result of this article is to show that the singular term gives rise to a non-analytic behavior of the spectrum for small values of the noncommutativity parameter $\theta$. Indeed, we will prove that some of the eigenvalues show a $\log{\theta}$ behavior for small $\theta$.

\smallskip

In the commutative case a singular potential of the type $\alpha/r ^{2} $ leads to a formally scale invariant Schr\"odinger equation \cite{DeAlfaro}, which is closely related to the dynamics of quantum particles in the asymptotic near-horizon region of black-holes \cite{Gibbons,Moretti}, of particles in a manifold with a conic singularity\cite{Klaus} or in the presence of an Aharonov-Bohm magnetic flux \cite{Mikhail}.

We will study the following Hamiltonian, that contains the noncommutative generalization of the inverse square potential:
\begin{eqnarray}\label{b1}
H&=&\frac{1}{2m}\,P^2+\frac{m\omega^2}{2}R\,^2+\frac{\alpha}{R^2}\,.
\end{eqnarray}
As we will see, the operator $P^2$ can be regarded as the usual Laplacian; $m,\omega, \alpha^{-1}$ are positive parameters with mass dimension. $R^2$ is the squared distance to the origin in the non-commutative plane
\begin{equation}\label{rsqua}
    R^2:=X^i X^i
\end{equation}
where $X^{i=1,2}$ are non-commutative coordinates satisfying the algebra
\begin{equation}\label{alg}
    [X^i, X^j]=i\,\epsilon^{ij}\,\theta\,.
\end{equation}

Since we are interested in studying the spectrum of (\ref{b1}) perturbatively in $\alpha$, we will first consider in Section \ref{conmu} the validity of this kind of perturbative treatment of this problem in the usual commutative plane. In Section \ref{pertu} we will state with no proof the theorems we will use in Section \ref{noconmu} to justify a perturbative treatment of the noncommutative case. For completeness, this theorems will also be applied to the usual commutative case in the Appendix \ref{percon}. In Section \ref{nonanal} we compute the spectrum to leading order in $\alpha$ and show that the energies of the noncommutative Hamiltonian (\ref{b1}) are not analytic in the noncommutativity parameter $\theta$. In Section \ref{conclu} we draw our conclusions.

\section{Commutative Plane}\label{conmu}

In this section we determine the spectrum of a non-relativistic particle living on the usual commutative plane subject to a central potential which consists in a singular term  $\alpha/r^2$ and a term proportional to $r^2$, where $r$ is the distance to the origin. First, we present the exact solutions and then we study the validity of perturbation theory in the parameter $\alpha$. The corresponding Hamiltonian is given by
\begin{equation}\label{b2}
H=\frac{1}{2m}\,p^2+\frac{m\omega^2}{2}\,r^2+\frac{\alpha}{r^2}
\end{equation}
Analogous operators appearing in one and two-dimensional problems have already been studied in \cite{FP1,FPW}. Since the Hamiltonian commutes with the angular momentum operator $L=-i\partial_\varphi$ we restrict the eigenvalue equation
\begin{equation}\label{eigequ}
(H-E)\,\Psi(r,\varphi)=0
\end{equation}
to the invariant subspace of functions
\begin{equation}\label{angmom}
\Psi(r,\varphi)= e^{i\varphi l}\psi_{l}(r)
\end{equation}
where $l\in\mathbb{Z}$. In the subspace of vanishing angular momentum, the operator (\ref{b2}) admits a one-parameter family of self-adjoint extensions \cite{RS2} which, in general, break the formal scale invariance of the Hamiltonian (near the origin) \cite{FP1,FPW,FMP,FMPS}. As we will see, it is convenient to consider the cases $l\neq 0$ and $l=0$ separately.

\subsection*{Subspaces with non-vanishing angular momentum}\label{SCNC}

In this subspaces, eq.\ (\ref{eigequ}) reads
\begin{equation}\label{b4}
\left(\frac{d^2}{dr^2}\,+\frac{1}{r}\frac{d}{dr}\,-\frac{\nu^2}{r^2}\,-\beta^2\,r^2 \,+ \lambda\right)\psi_{l}(r)=0\,,
\end{equation}
where
\begin{eqnarray}\label{b8}
\nu&:=&\mbox{}+\sqrt{l^2+2m\alpha}\,,\nonumber \\
\beta&:=&m\omega\,,\\
\lambda&:=&2mE\,.\nonumber
\end{eqnarray}
Since we are considering $\alpha\geq 0$, then $\nu\geq 1$. Eq.\ (\ref{b4}) has the following complete set of solutions $\psi_{l,\,n}(r)\in L_2(\mathbb{R}^+,r\,dr)$
\begin{equation}\label{b7-repetido}
\psi_{l,\,n}(r)=C_{l,n}\, r^{\nu}e^{-\frac{\beta}{2}r^2}M\left(-n,1+\nu;\beta r^2\right)
\end{equation}
where $M\left(-n,1+\nu;\beta r^2\right)$ is a confluent hypergeometric function \cite{A-S}, $C_{l,n}\in\mathbb{C}$ is a normalization constant and $n=0,1,2,\ldots$

The corresponding eigenvalues of eq.\ (\ref{eigequ}) are given by
\begin{eqnarray}\label{b9}
    E_{l,\,n}=\omega\,\left(2n+1 + \sqrt{l^2+2m\alpha}\right)
\end{eqnarray}
with $n=0,1,2,\ldots$ and $l=\pm1,\pm2,\pm3,\ldots$ Notice that, since $l\neq 0$, each eigenvalue is, at least, two-fold degenerate.

Next, we show that this result can be obtained perturbatively to leading order in $\alpha$. We begin by considering the spectrum of the unperturbed Hamiltonian $H_0$, which corresponds to an isotropic harmonic oscillator on the plane,
\begin{equation}\label{b2sp}
H_0=\frac{1}{2m}\,p^2+\frac{m\omega^2}{2}\,r^2\,.
\end{equation}
Its eigenfunctions $\Psi^{(0)}_{n,l}(r,\varphi)$ and eigenvalues $E^{(0)}_{n,l}$ --which can be obtained from (\ref{angmom}), (\ref{b7-repetido}) and (\ref{b9}) by setting $\alpha=0$-- are given by
\begin{eqnarray}
\Psi^{(0)}_{l,\,n}(r,\varphi)&=&C^{(0)}_{l,n}\ r^{|l|}e^{-\frac{\beta}{2}r^2}M\left(-n,1+|l|,\beta r^2\right)\,e^{i\varphi l}\,,\label{b9spp}\\
E^{(0)}_{l,\,n}&=&\omega\,\left(2n+1 + |l|\right)\,,\label{b9sp}\
\end{eqnarray}
where
\begin{equation}\label{con}
C^{(0)}_{l,n}= \sqrt{\frac{\beta^{|l|+1}}{\pi}\left(\begin{array}{c} n+|l|\\|l|\end{array} \right)}\,,
\end{equation}
with $n=0,1,2,\ldots$ and $l=\pm1,\pm2,\pm3,\ldots$

The leading order corrections $\Delta E^{(0)}_{l,\,n}$ to the energies $E^{(0)}_{l,\,n}$ due to singular term $\alpha/r^2$ are given by
\begin{equation}
    \Delta E^{(0)}_{l,\,n}=\left< \Psi^{(0)}_{l,\,n}
    \left|\frac{\alpha}{r^2}\right| \Psi^{(0)}_{l,\,n} \right>
    =\frac{m\omega\alpha}{|l|}\,.
\end{equation}
Therefore, a perturbative calculation in $\alpha$ gives for the energies
\begin{equation}\label{b12}
    E_{l,\,n}\simeq E^{(0)}_{l,\,n}+\Delta E^{(0)}_{l,\,n}=\omega\left(2n+1+|l|+\frac{m\alpha}{|l|}\right)
\end{equation}
which, as expected, corresponds to the first term in the Taylor expansion in $\alpha$ of (\ref{b9}).

We see that, in the non-vanishing angular momentum subspaces, corrections to the energies due to the singular term $\alpha/r^2$ can be obtained by ordinary perturbation theory in $\alpha$. We give a complete proof of this result in Appendix \ref{percon}. Next, we show that a perturbative treatment is not valid for the subspace corresponding to vanishing angular momentum.

\subsection*{Subspace with vanishing angular momentum}\label{SCC}

In this subspace, the formal scale invariance of the Hamiltonian (\ref{b2}) close to the singularity is, in general, broken by the existence of an infinite family of self-adjoint extensions\cite{FP1,FMP}. Indeed, for $0\leq 2m\alpha<1$, self-adjointness of the Hamiltonian does not suffice to determine the behavior of the wave functions at the origin. One must impose one of infinitely many admissible boundary conditions at $r=0$ which, in general, introduce a parameter with mass dimension.

However, there exists a unique self-adjoint extension for which the wave functions are finite at the singularity. For the sake of simplicity and for the purposes of this article, we will just consider the spectrum of this particular self-adjoint extension, which is given by the following eigenfunctions and eigenvalues
\begin{eqnarray}\label{aut}
    \Psi_{0,\,n}(r,\varphi)&=&C_{0,\,n}\, e^{-\frac{\beta}{2}r^2}r^{\nu_0}M\left(-n,1+\nu_0,\beta r^2\right)\\
    E_{0,\,n}&=&\omega\,\left(2n+1 + \sqrt{2m\alpha}\right)\qquad n=0,1,2,\ldots
\end{eqnarray}
where $C_{0,\,n}\in\mathbb{C}$ and $\nu_0=\sqrt{2m\alpha}$.

Since the energies are non-analytic around $\alpha=0$, we expect perturbation theory to fail in this case --as opposed to the $l\neq 0$ case.-- Indeed, the first order correction to the energy of the unperturbed (i.e., $\alpha=0$) eigenfunction
\begin{equation}\label{autsp}
\Psi^{(0)}_{0,\,n}(r)=\sqrt{\frac{\beta}{\pi}}\, e^{-\frac{\beta}{2}r^2}M\left(-n,1,\beta r^2\right)
\end{equation}
would be given by
\begin{equation}
    \Delta E^{(0)}_{0,\,n}=\left\langle \Psi^{(0)}_{0,\,n}
    \left|\frac{\alpha}{r^2}\right| \Psi^{(0)}_{0,\,n} \right\rangle=
    2\pi\,\int_0^{\infty} \frac{\alpha}{r^2}\cdot\left|\Psi^{(0)}_{0,\,n}(r,\varphi)\right|^2
    \,r\,dr
\end{equation}
which diverges at the lower limit $r=0$.

The non-validity of perturbation theory for vanishing angular momentum will be discussed in more detail in Appendix \ref{percon}.

\section{Perturbation Theory}\label{pertu}

In the previous section we have considered a two-dimensional isotropic harmonic oscillator in the presence of a singular potential of the form $\alpha/r^2$ to test the validity of perturbation theory for small $\alpha$. We concluded that the perturbative calculation in $\alpha$ is justified only in the non-vanishing angular momentum subspaces. In the next section we will consider the equivalent problem in the non-commutative plane, where perturbation theory is fully applicable. To prove the validity of perturbation theory for the non-commutative case, we need the following theorems and definitions \cite{RS4}:

\smallskip

{\defn\label{defK} An operator-valued function $H(\alpha)$ on a complex domain $\mathcal{R}$ is called an {\bf analytic family in the sense of Kato} if and only if:
\begin{enumerate}
\item $\forall \alpha\in \mathcal{R}$, $H(\alpha)$ is closed and has a nonempty resolvent set $\rho(H(\alpha))$.
\item $\forall \alpha_0\in \mathcal{R}, \exists \lambda_0\in \rho(H(\alpha_0))$ so that $\lambda_0\in\rho(H(\alpha))$ and $(H(\alpha)-\lambda_0)^{-1}$ is an analytic operator-valued function of $\alpha$, for $\alpha$ near $\alpha_0$.
\end{enumerate}}

\smallskip

{\thm [Kato-Rellich theorem]\label{KR} Let $H(\alpha)$ be an analytic family in the sense of Kato. Then for each isolated non-degenerate eigenvalue $E^{(0)}$ of $H(\alpha_0)$ and $\alpha$ near $\alpha_0$, there exists a non-degenerate eigenvalue $E(\alpha)$ in the spectrum of $H(\alpha)$ which is close to $E^{(0)}$. Moreover, $E(\alpha)$ is an analytic function of $\alpha$, for $\alpha$ near $\alpha_0$.}

\smallskip

{\defn\label{defA} An operator-valued function $H(\alpha)$ on a complex domain $\mathcal{R}$ is called an {\bf analytic family of type (A)} if and only if:
\begin{enumerate}
\item $\forall \alpha\in \mathcal{R}$, $H(\alpha)$ is closed and has a nonempty resolvent set $\rho(H(\alpha))$.
\item The operator domain of $H(\alpha)$ is some set $\mathcal{D}$ independent of $\alpha$.
\item $\forall \psi\in\mathcal{D}$, $H(\alpha)\psi$ is a vector-valued analytic function of $\alpha$.
\end{enumerate}}

\smallskip

{\thm\label{tipoA} Let $H_0$ be a closed operator with nonempty resolvent set $\rho(H_0)$ and consider another operator  $V$. If the domain of $H_0$ is contained  in the domain of $V$ and if $V$ is $H_0$-bounded, that is:
\begin{enumerate}
\item $\mathcal{D}(H_0) \subset\mathcal{D}(V)$
\item For some $A,B\in\mathbb{C}$ and  $\forall \psi\in\mathcal{D}(H_0): \|V\psi\|\leq A\|\psi\|+B\|H_0\psi\|$
\end{enumerate}
then $H(\alpha):=H_0+\alpha V$ is an analytic family of type (A).}

\smallskip

{\thm\label{AesK} Let $H(\alpha):=H_0+\alpha V$ be an analytic family of type (A) in some region $\mathcal{\mathcal{R}}$. Then $H(\alpha)$ is an analytic family in the sense of Kato.}

\smallskip

In Appendix \ref{percon} we will apply these theorems to the commutative case to justify the results obtained in the previous section. In the next section we will use these theorems in the noncommutative case to prove that the spectrum of the  operator (\ref{b1}) is analytic in the parameter $\alpha$.

\section{Noncommutative Plane}\label{noconmu}

To analyze the spectrum of the Hamiltonian (\ref{b1}) we adopt the following realization of the deformed Heisenberg algebra as in (\ref{alg}),
\begin{eqnarray}\label{b13}
X_{i}&=&x_{i}-\epsilon_{ij}\frac{\theta}{2}\,p_{j}\,,\\
P_{i}&=&p_{i}\,,
\end{eqnarray}
where  $x_i$ and $p_i$ satisfy the canonical Heisenberg algebra on the Hilbert space $\mathbf{L}_2(\mathbb{R}^2)$. In this representation  $R^2$ (see eq.\ (\ref{rsqua})) is given by the following second-order differential operator
\begin{equation}\label{a0-repetido}
    R^2=r^2- {\theta}L+ \frac{\theta^2}{4}p^2\,.
\end{equation}
The Hamiltonian (\ref{b1}) can then be written as
\begin{equation}\label{h00}
    H=H_{0}+\frac{\alpha}{R^2}\,,
\end{equation}
where
\begin{equation}\label{h0}
H_0=
    \frac{1}{2\mu}\,p^2+\frac{\mu\Omega^2}{2}\,r^2 -
    \frac{\mu\Omega^2\theta}{2}\,L
\end{equation}
and
\begin{eqnarray}\label{masa}
    \mu&:=&\frac{m}{1+\frac{m^2\omega^2\theta^2}{4}}\,,\\\nonumber \\
    \Omega&:=& \sqrt{1+\frac{m^2\omega^2\theta^2}{4}}\ \omega \,.
\end{eqnarray}
Notice that the angular momentum operator $L=\epsilon_{ij}\,x_ip_j$ in eq.\ (\ref{h0}) is the usual one in the commutative space.

Unlike the Hamiltonian for the commutative case studied in Section \ref{conmu}, the spectrum of (\ref{b1}) can be determined perturbatively in $\alpha$ for all values of the angular momentum, even for $l=0$. Let us next justify this statement using the theorems of Section \ref{pertu}.

Let us first consider the unperturbed Hamiltonian, i.e., the rotationally invariant operator $H_0$ (see eq.\ (\ref{h0}).) Notice that its restriction to the subspace of angular momentum $l$, $H_0^{(l)}$, corresponds --after the replacements $m\rightarrow\mu$, $\omega\rightarrow \Omega$ and $\alpha\rightarrow 0$-- to the Hamiltonian already given in eq.\ (\ref{b2}) restricted to the same subspace, minus the constant term $\mu\Omega^2\theta l/2$. As a consequence, from eqs.\ (\ref{b9spp}) and (\ref{b9sp}) we can read the eigenfunctions and eigenvalues of the unperturbed Hamiltonian $H_0$:
\begin{eqnarray}
\Psi^{(0)}_{l,\,n}(r,\varphi)&=& N^{(0)}_{l,\,n}\ r^{|l|}e^{-\frac{\mu\Omega}{2}r^2}M(-n,|l|+1,\mu\Omega r^2)\,e^{i\varphi l}\,, \label{b20}\\
E^{(0)}_{l,\,n}&=&\Omega\,\left(2n+1 + |l|-g\,l\right)\label{e20}\,,
\end{eqnarray}
where $l\in\mathbb{Z}$, $n=0,1,2,\ldots$,
\begin{equation}\label{g}
    g:=\frac{1}{2}\,\frac{m\omega\theta}{\sqrt{1+\left(\frac{1}{2}m\omega\theta\right)^2}}
    <1
\end{equation}
and
\begin{equation}
    N^{(0)}_{l,\,n}=\sqrt{\frac{(\mu\Omega)^{|l|+1}}{\pi}\left(\begin{array}{c} n+|l|\\|l|\end{array} \right)}\,.
\end{equation}

Our next purpose is to show that the energies of the Hamiltonian (\ref{b1}) on each angular momentum subspace are actually given --to leading order in $\alpha$-- by the quantities (\ref{e20}) plus the expectation values of the operator $\alpha/R^2$ with respect to the states (\ref{b20}).

The noncommutative squared distance to the origin $R^2$ is represented by a differential operator (eq.\ (\ref{a0-repetido})) whose spectrum can be obtained from that of $H_0$ (eq.\ (\ref{h0})) by making the following replacements: $\mu\rightarrow 2/\theta^2$ and $\Omega\rightarrow \theta$. Therefore, we can read from eqs.\ (\ref{b20}) and (\ref{e20}) the eigenfunctions $\Upsilon_{l,\,n}(r,\varphi)$ and eigenvalues  $\rho^2_{l\,,n}$ of the operator $R^2$:
\begin{eqnarray}\label{resdis}
\Upsilon_{l,\,n}(r,\varphi)&=& \sqrt{\frac{2^{|l|+1}(n+|l|)!}{\pi\,\theta^{|l|+1}n!|l|!}}\ r^{|l|}e^{-r^2/\theta}M(-n,|l|+1,2r^2/\theta)\
e^{i\varphi l}\,,
%\label{resdis}
\\
\rho^2_{l,\,n}&=&\theta\,(2n+1+|l|-l)\,,\label{resdis1}
\end{eqnarray}
with $n=0,1,2,\ldots$ and $l\in\mathbb{Z}$. Notice that, as a result of the non-commutativity of the coordinates, the distance to the origin is represented by an observable with a discrete set of eigenvalues. The minimum of the distance to the origin\footnote{This minimum is in accordance with the choice of the domain of the operator $R^2$, which in the present case corresponds to consider wave functions which are finite at the origin. Had we chosen a different set of functions, the minimum of the distance to the origin would have been different, even negative \cite{FMP}.} is $\sqrt{\theta}$.

On the other hand, in the subspace of angular momentum $l$ the perturbation $\alpha/R^2$ is proportional to the inverse of the differential operator $R^2$, whose kernel is a symmetric function, $G_l(r,r')=G_l(r',r)$, which, for $r<r'$, is given by
\begin{eqnarray}
    G_l(r,r')=\alpha\,\frac{2^{|l|+1}\Gamma(1/2+|l|/2-l/2)}{\theta^{|l|+2}\Gamma(|l|+1)}
    \ (rr')^{|l|}\,e^{-\frac{1}{\theta}(r^2+r'^2)}\times
    \\\times
    \ M(1/2+|l|/2-l/2,|l|+1,2r^2/\theta)\,U(1/2+|l|/2-l/2,|l|+1,2r'^2/\theta)\,,\nonumber
\end{eqnarray}
where $M(a,b,z)$ and $U(a,b,z)$ are two linearly independent solutions of the confluent hypergeometric equation \cite{A-S}.

In consequence, $\alpha/R^2$ is a (bounded) integral operator whose domain $\mathcal{D}(\alpha/R^2)$ is the whole Hilbert space $\mathbf{L}_2(\mathbb{R}^2)$. Thus, it contains the domain $\mathcal{D}(H_0)$ of $H_0$ and the first condition of Theorem \ref{tipoA} is satisfied for $V=\alpha/R^2$.

Moreover, since $\alpha/R^2$ is bounded, for all $\psi\in\mathbf{L}_2(\mathbb{R}^2)$ there exists a constant $C\in\mathbb{C}$ such that $\|\alpha/R^2\,\psi\|\leq C\|\psi\|$. Hence, the second condition of Theorem \ref{tipoA} is also satisfied and the Hamiltonian $H=H_0+\alpha/R^2$ (\ref{h00}) represents an analytic family of type (A) for any value of the parameter $\alpha$ in the region $R=\mathbb{C}$.

Theorem \ref{AesK} implies $H=H_0+\alpha/R^2$ is also analytic in the sense of Kato. Finally, Kato-Rellich Theorem \ref{KR} ensures that the spectrum of the Hamiltonian $H$ is analytic in $\alpha$, for $\alpha$ near zero. This allows us to perturbatively approach its eigenvalues as an expansion in integer powers of $\alpha$.

We will next determine the eigenvalues of $H$ by perturbing the eigenfunctions and spectrum of $H_0$, Eqs.\ (\ref{b20}) and (\ref{e20}), to leading order in $\alpha$. This calculation will show that the energies of the noncommutative-case Hamiltonian $H$ are not analytic functions of the non-commutativity parameter $\theta$.

\section{The perturbative spectrum}\label{nonanal}

In order to compute the perturbative corrections to the spectrum of $H_0$ due to the term $\alpha/R^2$ we use the spectral resolution (eqs.\ (\ref{resdis}) y (\ref{resdis1}))
\begin{equation}
    \frac{\alpha}{R^2}=\sum_{l'\in\mathbb{Z}}\sum_{n'=0}^\infty\frac{\alpha}{\rho^2_{l',\,n'}}
    \ |\Upsilon_{l',\,n'}\rangle \langle \Upsilon_{l',\,n'}|\,.
\end{equation}

The correction of first order in $\alpha$ to the energy $E^{(0)}_{l,\,n}$ in (\ref{e20}) is given by\footnote{We assume $\theta\neq 0$, though later we will study the limit $\theta\rightarrow 0$ of the resulting expressions for comparison with the commutative case.}
\begin{equation}
    \left\langle \Psi^{(0)}_{l,\,n}
    \left|\frac{\alpha}{R^2}\right| \Psi^{(0)}_{l,\,n} \right\rangle
    =\sum_{n'=0}^\infty
    \frac{\alpha}{\rho^2_{l,\,n'}}
    \ |\langle \Psi^{(0)}_{l,\,n}|\Upsilon_{l,\,n'}\rangle |^2
\end{equation}
which, after eqs.\ (\ref{b20}) and (\ref{resdis}), reads \cite{TabIntegrales}
\begin{eqnarray}
    \begin{array}{c}
    \displaystyle{
    \left\langle \Psi^{(0)}_{l,\,n}
    \left|\frac{\alpha}{R^2}\right| \Psi^{(0)}_{l,\,n} \right\rangle =}\displaystyle{
    \frac{2\alpha \mu\Omega\,(4g)^{|l|}}{(1+g)^{2|l|+2}}
    \frac{(n+|l|)!}{n!|l|!^4}
    \sum_{n'=0}^\infty
    \frac{(n'+|l|)!}{n'!(2n'+1+|l|-l)}\times}\\ \\
    \displaystyle{\times \left|\int_0^\infty dx\,e^{-x}x^{|l|}M\left(-n,1+|l|,\frac{2g}{1+g}x\right)
    M\left(-n',1+|l|,\frac{2}{1+g}x\right)\right|^2}\,\\ \\
    =  \displaystyle{\frac{2\alpha \mu\Omega\,(4g)^{|l|} (n+|l|)!}{(1+g)^{2|l|+2} n!|l|!^{2}} \left(\frac{1-g}{1+g}\right)^{2n}\times}\\ \\ \displaystyle{\times \sum_{n'=0}^{\infty}\left(\frac{(n'+|l|)!}{n'!(2n'+1+|l|-l)}\right)\left(\frac{1-g}{1+g}\right)^{2n'} \left[\mbox{}_2F_1\left(
    \left\{-n',-n\right\},
    \left\{1+|l|\right\};
    \frac{-4g}{(1-g)^2}
       \right)\right]}^2.\end{array}
\end{eqnarray}

As we already mentioned, we will show that this first order corrections are non-analytic in $\theta$ around $\theta=0$. In so doing, it will suffice to consider, for each angular momentum subspace, the perturbative correction to the lowest energy level $E^{(0)}_{l,\,n=0}$, given by
\begin{equation}\label{resu}
    \begin{array}{c}
    \displaystyle{
    \left\langle \Psi^{(0)}_{l,\,n=0}
    \left|\frac{\alpha}{R^2}\right| \Psi^{(0)}_{l,\,n=0} \right\rangle=}\displaystyle{
    \frac{2\alpha \mu\Omega\,(4g)^{|l|}}{(1+g)^{2|l|+2}}
    \,\times}\\ \\
    \displaystyle{\times\,
    \sum_{n'=0}^\infty
    \left(\begin{array}{c}n'+|l|\\ |l| \end{array}\right)
    \frac{1}{2n'+1+|l|-l}
    \ \left(\frac{1-g}{1+g}\right)^{2n'}
    }
    \\ \\
    =\displaystyle{
    \frac{2\alpha \mu\Omega\,(4g)^{|l|}}{(1+|l|-l)\,(1+g)^{2|l|+2}}\,\times}\\ \\
    \displaystyle{\times\,
    \mbox{}_2F_1\left(
    \left\{1+|l|,\frac{1}{2}(1+|l|-l)\right\},
    \left\{\frac{1}{2}(3+|l|-l)\right\};
    \left(
    \frac{1-g}{1+g}
    \right)^2
    \right)\,.
    }\end{array}
\end{equation}

This expression represents the leading correction to the lowest energy at each angular momentum subspace due to the term $\alpha/R^2$ for any value of the noncommutativity parameter (contained in) $g$ (see eq.\ \eqref{g}.) To show that this correction is non-analytic in $g$ --or, equivalently, in $\theta$-- we make use of the following expansion of the hypergeometric function
%$\mbox{}_2F_1\left(\left\{1+|l|,c\right\},\left\{1+c\right\};1-\epsilon\right)$
\cite{A-S}:
\begin{eqnarray}\label{hyp}
    \displaystyle{
    \mbox{}_2F_1\left(
    \left\{1+|l|,c\right\},
    \left\{1+c\right\};1-\epsilon
    \right)=
    \frac{c}{|l|}\, \epsilon^{-|l|}\,
    \sum_{n=0}^{|l|-1}\frac{(c-|l|)_n (1)_n}{n!(1-|l|)_n}\,\epsilon^n-}\,,\\\displaystyle{
    \mbox{}-(-1)^{|l|}\,\frac{\Gamma(c+1)}{\Gamma(c-|l|)}
    \sum_{n=0}^{\infty}\frac{(c)_n}{n!|l|!}\,\epsilon^n\ \left\{
    \log{\epsilon}+\psi(n+c)-\psi(n+1)
    \right\}}\,,\nonumber
\end{eqnarray}
where $0<\epsilon<1$ and $l\neq 0$. In (\ref{hyp}),  $\psi(\cdot)$ is the logarithmic derivative of the gamma-function $\Gamma(\cdot)$, and $(\cdot)_n$ represents the Pochhammer symbol\footnote{$(c)_n:=\Gamma(c+n)/\Gamma(c)$.}. Non-analyticity in $\theta$ is due to the $\epsilon^n\log\epsilon$ terms in the series on the right hand side of this equation.

Replacing (\ref{hyp}) into (\ref{resu}) and taking into account the eigenvalues of $H_0$ given by \eqref{e20} we obtain the lowest energies of the complete Hamiltonian $H$ for fixed angular momentum and small $\alpha$ and $\theta$,
\begin{equation}
\begin{array}{ll}
    \displaystyle{\frac{E_{l,\,n=0}}{\omega}=2+\alpha m\mp \frac{m\omega\theta}{2}\mp \alpha m^2\omega\theta
    \left\{\log{\left(\frac{m\omega\theta}{2}\right)}+1\right\}+\dots}
    &\displaystyle{ {\rm if}\ |l|=1}\,,\\ \\
    \displaystyle{\frac{E_{l,\,n=0}}{\omega}=1+|l|+\frac{\alpha m}{|l|}
    \mp |l|\,\frac{m\omega\theta}{2}\mp \frac{\alpha m^2\omega\theta}{|l|-1}
    +\dots}
    &\displaystyle{ {\rm if}\ |l|\geq  2}\,.
\end{array}
\end{equation}
The upper and lower signs correspond respectively to positive and negative angular momentum states, and reflect the time-reversal non-invariance of this system.

Notice that, in spite of the $\log{\theta}$ terms in the energies, the commutative limit $\theta\rightarrow 0$ exists and reproduces the result in (\ref{b12}), obtained in the previous section for the commutative-case Hamiltonian for non-vanishing angular momentum subspaces.

\smallskip

Finally, we consider the $l=0$ case; from eq.\ \eqref{resu} we simply get
\begin{equation}\label{resu-repetido}
    \begin{array}{rl}
    \displaystyle{\left\langle \Psi^{(0)}_{l=0,\,n=0}
    \left|\frac{\alpha}{R^2}\right| \Psi^{(0)}_{l=0,\,n=0} \right\rangle
    }&\displaystyle{=
    \frac{2\alpha \mu\Omega}{(1+g)^{2}}\ \cdot\ \mbox{}_2F_1\left(
    \left\{1,\frac{1}{2}\right\},
    \left\{\frac{3}{2}\right\};
    \left(
    \frac{1-g}{1+g}
    \right)^2
    \right)}\\ \\ &\displaystyle{
    =
    -\,\frac{\alpha \mu\Omega}{1-g^2}\,  \log{g}
    }\,.
    \end{array}
\end{equation}
Combining Eqs.\ \eqref{resu-repetido} and \eqref{e20} we obtain the ground state energy to leading order in $\alpha$ for arbitrary values of the noncommutativity parameter $\theta$,
\begin{equation}\label{e20-repetido}
    \frac{E_{l=0,\,n=0}}{\omega}=
    \sqrt{1+\left(\frac{m\omega\theta}{2}\right)^2}\left\{
    1-\alpha m\ \log{\left(\frac{\frac{m\omega\theta}{2}}
    {\sqrt{1+\left(\frac{m\omega\theta}{2}\right)^2}}
    \right)}\right\}+O(\alpha^2)\,.
\end{equation}

Expression (\ref{e20-repetido}) shows that the correction for non-commutativity of the space to the ground state energy is non-analytic in $\theta$ around $\theta=0$, being proportional to $\log\theta$ for small $\theta$. Moreover, this correction diverges in the $\theta\rightarrow 0$ limit. This is consistent with the fact already shown in section \ref{conmu} that perturbation theory in $\alpha$ is not applicable on the vanishing angular momentum subspace for the commutative case.

\section{Conclusions}\label{conclu}

The generalization of Schr\"{o}dinger operators to the noncommutative plane can be carried out by means of the Bopp shift $X_{i}=x_{i}-\epsilon_{ij}\theta\,p_{j}/2$. Since the noncommutative coordinates are then represented by differential operators, the subsequent generalization of the Schr\"{o}dinger Hamiltonian $-\triangle + V(X)$ changes drastically its character as a linear operator on the Hilbert space. In particular, positive powers $X^n$ of the coordinates in the potential $V(X)$ correspond to differential operators of order $n$ whereas an inverse square $1/R^{2}$, as in the case of the present article, renders a non-local bounded operator.

As a consequence, the spectral properties of this kind of Hamiltonians are different from the well-known properties of second order Schr\"{o}dinger operators. In this article we gave an example of how analyticity in one of the parameters of the Hamiltonian, namely the noncommutativity parameter $\theta$, is spoiled as a consequence of this non-local representation of the Schr\"{o}dinger potential. This result is to be remarked since, due to the non-standard character of noncommutative-case Hamiltonians, many of the known studies of problems in noncommutative spaces rely on expansions in powers of the parameter $\theta$.

In particular, we have studied the case of a singular potential and showed that the spectrum is not analytic in the noncommutativity parameter. We have considered on the noncommutative plane a perturbation proportional to the inverse squared distance to the origin, $\alpha/R^2$. Since this perturbation is represented by a bounded operator, the spectrum can be computed perturbatively in $\alpha$ and the non-analyticity of the energies in $\theta$ emerges explicitly as logarithmic cuts in $\theta=0$.

\bigskip

\noindent{\textbf{Acknowledgements}}:  This  work was  partially supported by CONICET (PIP 01787), ANPCyT (PICT 00909) and UNLP (Proy.~11/X492), Argentina. M.N. also acknowledge support from Universidad Nacional de La Plata, Argentina.

\medskip

\appendix

\section{Kato-Rellich theorem applied to the commutative case}\label{percon}

We have shown in Section \ref{conmu} that in the usual commutative case the spectrum is analytic in the parameter $\alpha$ if and only if $l\neq 0$ (see eqs. (\ref{b9}) and (\ref{aut})). In this appendix we will justify this results by means of the theorems stated in Section \ref{pertu}. In so doing, we will consider the cases $|l|\geq 2$ and $|l|=1$ separately.

\subsection{Subspaces with $|l|\geq 2$}

We will prove that, on this subspaces, the operator $H_0^{(l)}+\alpha/r^2$ --being $H_0^{(l)}$ the restriction of (\ref{b2sp}) to a fixed angular momentum subspace-- is an analytic family of type (A), for $\alpha$ in some region $\mathcal{R}$ around the origin.

Let us prove the first condition of Theorem \ref{tipoA}, namely, $\mathcal{D}(H^{(l)}_0) \subset\mathcal{D}(\alpha/r^2)$. To do that, consider the operator
\begin{equation}
    H_0^{(l)}=\frac{1}{2m}\left\{-\partial_{r}^2-\frac{\partial_{r}}{r}+\frac{l^2}{r^2}\right\}+\frac{m\omega^2}{2} \,r^2\,.
\end{equation}
If $\phi(r)\in\mathcal{D}(H_0^{(l)})$ then $H_0^{(l)}\phi(r)\in\mathbf{L}_2(\mathbb{R}^+,r\,dr)$. Since $r^2\phi(r)\in\mathbf{L}_2((0,1),r\,dr)$ then
\begin{equation}\label{apf}
    f(r):=\left(\partial_{r}^2+\frac{\partial_{r}}{r}-\frac{l^2}{r^2}\right)\phi(r)
    \in\mathbf{L}_2((0,1),r\,dr)\,.
\end{equation}
For $r\neq 0$ we define
\begin{equation}\label{uphi}
    u(r):=r^{|l|}\,\phi
\end{equation}
which, due to eq.\ (\ref{apf}), satisfies
\begin{equation}
    \partial_r(r^{-2|l|+1}\partial_r u(r))=r^{-|l|+1}\,f(r)\,.
\end{equation}
Therefore,
\begin{equation}\label{apu}
    \partial_r u(r)=2|l|C\,r^{2|l|-1}+r^{|l|}\,v(r)
\end{equation}
for some constant $C\in\mathbb{C}$, where
\begin{equation}
    v(r):=-r^{|l|-1}\int_r^1 x^{-|l|+1}f(x)\,dx\,.
\end{equation}
This function is well-defined since $r^{-|l|}$ and $f(r)$ belong to $\mathbf{L}_2((\epsilon,1),r\,dr)$, for any $\epsilon>0$. Moreover, $v(r)$ is continuous and finite at $r=0$.

Similarly, one gets from eqs.\ (\ref{uphi}) and (\ref{apu})
\begin{equation}\label{phior}
    \phi(r)=C\,r^{|l|}+w(r)\,,
\end{equation}
where
\begin{equation}\label{casi}
\begin{array}{rl}
    w(r):=&\displaystyle{r^{-|l|}\int_0^r x^{|l|}v(x)\,dx}\\ \\
    =&\displaystyle{-r^{-|l|}\int_0^r dx\,\int_x^1 dy\ x^{2|l|-1} y^{-|l|+1}f(y)}\\ \\
    =&\displaystyle{-\frac{1}{2|l|}
    \left[r^{-|l|}\,\int_0^r dy\ y^{|l|+1}f(y)+r^{|l|}
    \int_r^1 dy\ y^{-|l|+1}f(y)\right]}\,.
\end{array}
\end{equation}
By the Cauchy-Schwarz inequality one can prove that the first term in this last expression is $o(r)$. As regards the second term, we can write
\begin{equation}\label{int}
    r^{|l|}
    \int_r^1 dy\ y^{-|l|+1}f(y)=r\, \left(\chi_r(y),f(y)\right)\,,
\end{equation}
where the scalar product is taken in the space $\mathbf{L}_2((0,1),y\,dy)$ and
\begin{equation}
    \chi_r(y):=\left\{\begin{array}{cc}
    r^{|l|-1}\,y^{-|l|}& {\rm if\ }r<y<1\\
    0& {\rm if\ }0<y<r
    \end{array}\right.\,.
\end{equation}
These functions converge uniformly to zero on every compact interval not containing the origin. Therefore, the scalar product $(\chi_r(y),\varphi(y))\rightarrow 0$ as $r\rightarrow 0$ for all $\varphi\in\mathcal{C}_0^{\infty}(0,1)$ --the space of smooth functions with compact support in $(0,1)$.--

On the other hand, since $\mathcal{C}_0^{\infty}(0,1)$ is dense in $\mathbf{L}_2((0,1),y\,dy)$, for each $f\in\mathbf{L}_2((0,1),y\,dy)$ there exists a sequence $\varphi_n\rightarrow f$. As a consequence, the scalar product $(\chi_r(y),f(y))\rightarrow 0$ as $r\rightarrow 0$ for all $f\in\mathbf{L}_2((0,1),y\,dy)$. Indeed,
\begin{equation}
    |(\chi_r,f)| \leq |(\chi_r,f-\varphi_n)|+|(\chi_r,\varphi_n)|
\end{equation}
and the \small{R.H.S} of this expression can be made arbitrarily small; the first term for $n$ large enough, the second one for $r$ small enough. We conclude that expression (\ref{int})) and, consequently, the second term in the last expression of eq.\ (\ref{casi}) is also $o(r)$.

Thus (see eq.\ (\ref{phior})) the functions $\phi(r)\in\mathcal{D}(H_0^{(l)})$ satisfy $\phi(r)=o(r)$. This is a sufficient condition to assert that $\phi\in\mathcal{D}(\alpha/r^2)$. This proves the first condition of Theorem \ref{tipoA} .

Next, we prove the second condition by computing
\begin{equation}
    \begin{array}{c}\displaystyle
    \|H_0^{(l)}\phi(r)\|^2=\frac{1}{4m^2}\int_0^\infty dr\,r\left|\left\{
    -\partial_{r}^2-\frac{\partial_{r}}{r}+\frac{l^2}{r^2}+m^2\omega^2r^2
    \right\}\phi(r)\right|^2
    \\ \\ \displaystyle
    =\frac{1}{4m^2}\int_0^\infty dr\,r\left\{
    \left|\frac{1}{r}\partial_r\left[r\partial_r \phi(r)\right]\right|^2+
    2\left(\frac{|l|^2}{r^2}+m^2\omega^2r^2\right)\left|\partial_r\phi(r)\right|^2+\right.
    \\ \\ \displaystyle
    \left.
    \mbox{}+\left(2m^2\omega^2(|l|^2-2)+m^4\omega^4r^4
    +\frac{|l|^2(|l|^2-4)}{r^4}\right)\left|\phi(r)\right|^2
    \right\}
    \\ \\ \displaystyle
    \geq\int_0^\infty dr\,r\ \frac{|l|^2(|l|^2-4)}{r^4}\left|\phi(r)\right|^2\,,
\end{array}
\end{equation}
which shows that
\begin{equation}
    \left\|\frac{\alpha}{r^2}\,\phi(r)\right\|\ \leq B\,\|H_0^{(l)}\,\phi(r)\|
\end{equation}
for some constant $B\in\mathbb{C}$. This proves the second condition of Theorem \ref{tipoA}. In consequence, $H_0^{(l)}+\alpha/r^2$ is an analytic family of type (A). Therefore it is also an analytic family in the sense of Kato (Theorem \ref{AesK}) and then, by virtue of Kato-Rellich theorem (Theorem \ref{KR}), its spectrum is analytic in $\alpha$.

\subsection{Subspaces with $|l|=1$}

For this subspaces the situation is slightly different. The operators $H_0^{(\pm 1)}+\alpha/r^2$ --where $H_0^{(\pm 1)}$ is the restriction of the operator given in eq.\ (\ref{b2sp}) to the subspaces with $l=\pm 1$ respectively-- is not an analytic family of type (A). As we have already mentioned in the previous section, the domain of the perturbation $\mathcal{D}(\alpha/r^2)$ contains functions which behave as $o(r)$ for small $r$. However, the domains of $H_0^{(\pm 1)}+\alpha/r^2$ contain functions which behave as $O(r^{\sqrt{1+2 m \alpha}})$ (see e.g., the eigenfunctions (\ref{b7-repetido}) for $l=\pm 1$.) As a consequence, there is no region $\mathcal{R}$ in the complex plane containing the origin for which there exists a domain $\mathcal{D}(H_0^{(\pm 1)}+\alpha/r^2)$ independent of $\alpha$. Since the second condition of Definition \ref{defA} is not fulfilled, $H_0^{(\pm 1)}+\alpha/r^2$ is not an analytic family of type (A).

Nevertheless, one can prove that $H_0^{(\pm 1)}+\alpha/r^2$ is actually analytic in the sense of Kato. Indeed, since $H_0^{(\pm 1)}+\alpha/r^2$ is selfadjoint it is a closed operator and its spectrum is real, then every non-real $\lambda$ belongs to its resolvent set. Therefore, the first condition in Definition \ref{defK} is satisfied.

As regards the second condition of Definition \ref{defK}, the resolvent $(H_0^{(\pm 1)}+\alpha/r^2-\lambda_0)^{-1}$ is an integral operator with a symmetric kernel $G(r,r',\lambda_0)=G(r',r,\lambda_0)$ that, for $r<r'$, reads
\begin{eqnarray}\label{myu}
    \mbox{}\\
    G(r,r',\lambda_0)=\omega^{\nu_1}m^{\nu_1+1}
    \ \frac{\Gamma(\nu_1/2+1/2-\lambda_0/2\omega)}{\Gamma(\nu_1+1)}
    (rr')^{\nu_1}\ e^{-\frac{m\omega}{2}(r^2+r'^2)}\times\nonumber\\
    \times\ M\left(\frac{\nu_1+1}{2}-\frac{\lambda_0}{2\omega},\nu_1+1, m\omega r^2\right)
    \,U\left(\frac{\nu_1+1}{2}-\frac{\lambda_0}{2\omega},\nu_1+1, m\omega r'^2\right)\nonumber
\end{eqnarray}
with
\begin{equation}
    \nu_1:=\sqrt{1+2m\alpha}\,.
\end{equation}
To show that the confluent hypergeometric functions $M$ and $U$ in eq.\ (\ref{myu}) are analytic in some neighbourhood around $\nu=1$ we use the following integral representations \cite{A-S},
\begin{eqnarray}
    M\left(\frac{\nu_1+1}{2}-\frac{\lambda_0}{2\omega},\nu_1+1, z\right)&=&
    \frac{\Gamma\left(\nu_1+1\right)}
    {\Gamma\left(\frac{\nu_1+1}{2}-\frac{\lambda_0}{2\omega}\right)
    \Gamma\left(\frac{\nu_1+1}{2}+\frac{\lambda_0}{2\omega}\right)}\times\label{m}\\
    &&\mbox{}\times\int_0^1 t^{\frac{\nu_1-1}{2}-\frac{\lambda_0}{2\omega}}
    \ (1-t)^{\frac{\nu_1-1}{2}+\frac{\lambda_0}{2\omega}}
    \ e^{zt}\,dt  \,,\nonumber \\ \nonumber \\
    U\left(\frac{\nu_1+1}{2}-\frac{\lambda_0}{2\omega},\nu_1+1, z\right)&=&
    \frac{1}
    {\Gamma\left(\frac{\nu_1+1}{2}-\frac{\lambda_0}{2\omega}\right)
    }\times\label{u}\\
    &&\mbox{}\times\int_0^\infty t^{\frac{\nu_1-1}{2}-\frac{\lambda_0}{2\omega}}
    \ (1+t)^{\frac{\nu_1-1}{2}+\frac{\lambda_0}{2\omega}}
    \ e^{-zt}\,dt\,, \nonumber
\end{eqnarray}
valid for $-\omega(\nu_1+1)<\mathcal{R}(\lambda_0)$. If we consider $\lambda_0$ such that $-\omega(\nu_1-1)<\mathcal{R}(\lambda_0)<\omega(\nu_1+1)$, the integral in eq.\ (\ref{m}) can be written as
\begin{equation}\label{f}
    \int_0^1
    f(\nu_1,t)\,dt
    +\frac{1}{\frac{\nu_1+1}{2}-\frac{\lambda_0}{2\omega}}
\end{equation}
where
\begin{equation}
    f(\nu_1,t):=t^{\frac{\nu_1-1}{2}-\frac{\lambda_0}{2\omega}}
    \ \left[(1-t)^{\frac{\nu_1-1}{2}+\frac{\lambda_0}{2\omega}}
    \ e^{zt}-1\right]\,.
\end{equation}

On the other hand, for $\mathcal{R}(\lambda_0)<\omega(\nu_1+1)$, the integral in eq.\ (\ref{u}) can be written as
\begin{equation}\label{inte}
    \int_0^1 g(\nu_1,t)\,dt
    +\frac{1}{\frac{\nu_1+1}{2}-\frac{\lambda_0}{2\omega}}+\int_1^\infty h(\nu_1,t)\,dt
\end{equation}
where
\begin{eqnarray}
    g(\nu_1,t)&:=&t^{\frac{\nu_1-1}{2}-\frac{\lambda_0}{2\omega}}
    \ \left[(1+t)^{\frac{\nu_1-1}{2}+\frac{\lambda_0}{2\omega}}
    \ e^{-zt}-1\right]\,, \\
    h(\nu_1,t)&:=&t^{\frac{\nu_1-1}{2}-\frac{\lambda_0}{2\omega}}
    \ (1+t)^{\frac{\nu_1-1}{2}+\frac{\lambda_0}{2\omega}}
    \ e^{-zt}\,.
\end{eqnarray}
Since there exists some $\epsilon_1>0$ such that: (a) $f(\nu_1,t),g(\nu_1,t)$ and their first derivatives with respect to $\nu_1$ are continuous on $(\nu_1,t)\in \{\nu_1:|\nu_1-1|<\epsilon_1\}\times[0,1]$ and (b) for all $\nu_1\in \{\nu_1:|\nu_1-1|<\epsilon_1\}$, $f(\nu_1,t)$ and $g(\nu_1,t)$ are analytic for all $t\in[0,1]$, then the integral in eq.\ (\ref{f}) and the integral in the first term of eq.\ (\ref{inte}) are analytic in the vicinity of $\nu_1=1$ \cite{Shilov}.

Finally, there exists some $\epsilon_2>0$ such that: (a) $h(\nu_1,t)$ and its first derivative with respect to $\nu_1$ are continuous on $(\nu_1,t)\in \{\nu_1:|\nu_1-1|<\epsilon_2\}\times[1,T)$, for all $1<T<\infty$ {and $\int_1^\infty h(\nu_1,t)\,dt$ converges uniformly on $|\nu_1-1|<\epsilon_2$} and (b) for all $\nu_1\in \{\nu_1:|\nu_1-1|<\epsilon_2\}$, $g(\nu_1,t)$ is analytic for all $t\in[1,\infty)$ and the integral $\int_1^\infty f(\nu_1,t)\,dt$ converges uniformly. Therefore, the integral in the third term of eq.\ (\ref{inte}) is also analytic in a neighborhood of $\nu_1=1$ \cite{Shilov}.

This proves that, for $-\omega(\nu_1-1)<\mathcal{R}(\lambda_0)<\omega(\nu_1+1)$, the kernel in (\ref{myu}) is analytic in some neighborhood of $\nu_1=1$. This is consistent with the fact that the spectra of the operators $H_0^{(\pm 1)}+\alpha/r^2$  are analytic in $\alpha$, as explicitly shown in eq.\ (\ref{b9}).

\end{document}